\documentclass[12pt,preprint]{aastex}
\slugcomment{accepted for publication Astrophysical Journal Letters}
\shorttitle{Discovery of an \ion{O}{6} Emitting Nebula}
\shortauthors{B. Otte et al.}

\begin{document}
\title{Discovery of an \ion{O}{6} Emitting Nebula around the Hot White Dwarf
KPD~0005+5106}
\author{Birgit Otte, W. Van Dyke Dixon, Ravi Sankrit}
\affil{Department of Physics and Astronomy, The Johns Hopkins University, 3400
North Charles Street, Baltimore, MD 21218}
\email{otte@pha.jhu.edu, wvd@pha.jhu.edu, ravi@pha.jhu.edu}

\clearpage

\begin{abstract}
A survey of diffuse interstellar sight lines observed with the {\em Far
Ultraviolet Spectroscopic Explorer} has led to the serendipitous discovery of a
high-ionization nebula around the hot white dwarf KPD~0005+5106. The nebula has
an \ion{O}{6} $\lambda$1032 surface brightness of up to 25,000
photons~s$^{-1}$~cm$^{-2}$~sr$^{-1}$, making it the brightest region of extended
\ion{O}{6} emission in our survey. Photoionization models using the incident
white dwarf continuum successfully reproduce the observed \ion{O}{6} intensity.
The \ion{O}{6} emission arises in the highly ionized inner region of a planetary
nebula around KPD~0005+5106. This newly discovered nebula may be one member of a
class of high-ionization planetary nebulae that are difficult to detect in the
optical, but which can be easily identified in the ultraviolet.
\end{abstract}



\section{INTRODUCTION}

Planetary nebulae (PNe) come in a variety of shapes, irregular, bipolar, or
spherical, and exhibit emission from forbidden lines such as [\ion{O}{3}],
[\ion{Ne}{3}], and [\ion{Ne}{5}] as well as from the H$\alpha$ recombination
line. In addition, numerous high-excitation PNe exhibit most of the other Balmer
lines as well as \ion{He}{2}, \ion{C}{3}], and other forbidden lines such as
[\ion{Ne}{4}], [\ion{Ar}{4}], [\ion{Mg}{5}]. The typical radial extent is a few
tenths of a parsec. Only a few PNe have radii exceeding 1.5~pc. Because of their
small size and low surface brightness, PNe are notoriously difficult to discover
\citep[e.g.,][]{kw,w97,rau}.

We report the discovery of an \ion{O}{6} emitting nebula around KPD~0005+5106
with the {\em Far Ultraviolet Spectroscopic Explorer (FUSE)} that we identify as
a high-ionization PN (HIPN) around this white dwarf (WD). This would make
KPD~0005+5106 only the third WD \citep[out of 162 WDs listed by][]{nap} known to
have a PN. Including KPD~0005+5106, only eight of the fifteen stars listed with
$T_{\rm eff}\geq120,000$~K \citep{nap} have a known PN.

\section{OBSERVATIONS}

The {\em FUSE} instrument consists of four optical systems. Two employ LiF
coatings and are sensitive to the wavelength range $990-1187$~\AA. The other two
use SiC coatings and are sensitive to wavelengths between 905 and 1100~\AA. Each
spectrograph possesses three apertures that simultaneously observe different
positions on the sky separated by a few arcminutes \citep[for a complete
description of {\em FUSE}, see][]{moos,sahn}.

\begin{figure}
\plotone{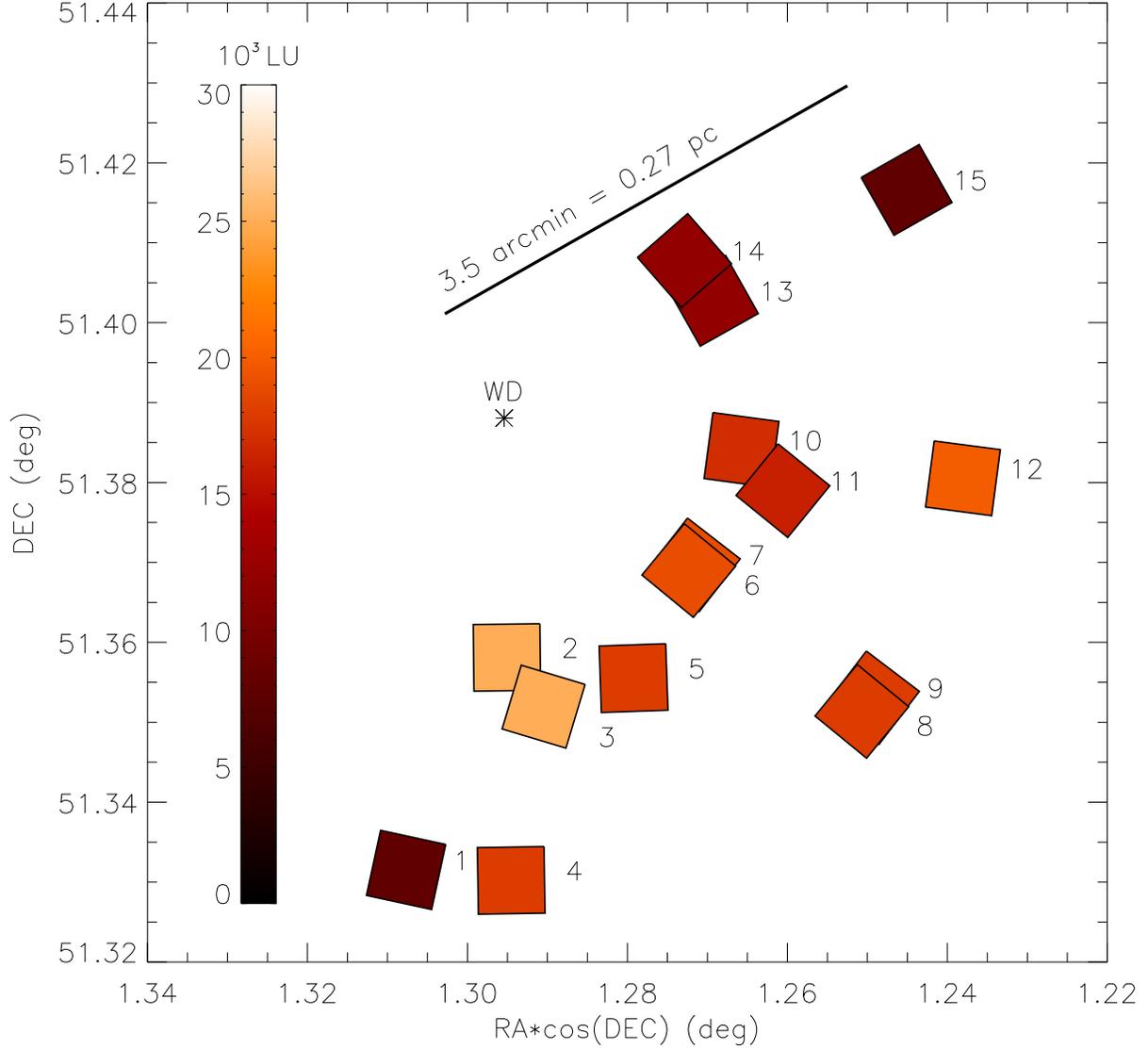}
\caption{\label{map} Observations of diffuse \ion{O}{6} emission around the hot
WD KPD0005+5106. Colors represent the \ion{O}{6} $\lambda$1032 intensity in
units of 10$^3$ LU (LU $\equiv$ photons~s$^{-1}$~cm$^{-2}$~sr$^{-1}$) recorded
through the {\em FUSE} LWRS aperture. The angular distance from the star to the
center of each aperture in the outer ring is 3.5 arcmin; at a distance of
270~pc, this corresponds to about 0.27~pc. Numbers refer to the map key listed
in Tables \ref{obs} and \ref {meas}.}
\end{figure}

Because of the strong interstellar H$_2$ features in its spectrum, the WD
KPD~0005+5106 was selected as a wavelength-calibration target and has been
observed repeatedly throughout the mission. When the {\em FUSE} high- or
medium-resolution aperture is centered on the star, the low-resolution (LWRS)
aperture samples a $30\arcsec\times30\arcsec$ region 1.8 or 3.5 arcmin
(respectively) away. Changes in the spacecraft roll angle from one observation
to another place the LWRS aperture at different position angles around the star.
Figure \ref{map} shows the observed LWRS positions. In three additional
background observations, the LWRS aperture was 2.2 arcmin from the star.

Table \ref{obs} lists the observational data for each pointing: observation ID,
equatorial coordinates of the LWRS aperture, date of the observation, and the
polar coordinates for the LWRS aperture relative to KPD~0005+5106. Angle
$\varphi$ is measured counterclockwise from the north celestial pole. A map key
identifies the LWRS positions in Figure \ref{map}. All observations were
performed in time-tag mode, in which the spectrograph records the position and
arrival time of each photon.

\section{DATA REDUCTION}

The individual exposures of each pointing were aligned using the measured
positions of the \ion{O}{1} airglow lines between 1027 and 1042~\AA. The data
were screened for pulse heights in the range $2-25$, shifted to a heliocentric
wavelength scale, and combined, and the region around the \ion{O}{6}
$\lambda\lambda$ 1032,1038 doublet was extracted. The spectra were binned by 16
pixels to increase the signal-to-noise ratio. A more detailed description of the
data reduction can be found in our \ion{O}{6} emission survey paper (Otte,
Dixon, \& Sankrit 2004, in preparation).

In most cases, we used night-only data to avoid possible contamination by an
unidentified airglow feature near the \ion{O}{6} $\lambda1032$ line. For
M1070208, M1070223, and M1070224, we combined day and night data to obtain a
more reliable measurement of the faint \ion{O}{6} emission; contamination by the
airglow feature could be excluded in these cases (by comparing the
day-plus-night and night-only data).

Due to their nearly identical positions, the data of observations M1070211 and
M1070214 were combined into one spectrum, as were the data of observations
M1070212 and M1070215. Although the overlap of their positions is less than for
the other two combined observations, we combined M1070209 and M1070221 to gain
signal-to-noise at the expense of spatial resolution.

\section{RESULTS}

\begin{figure}
\plotone{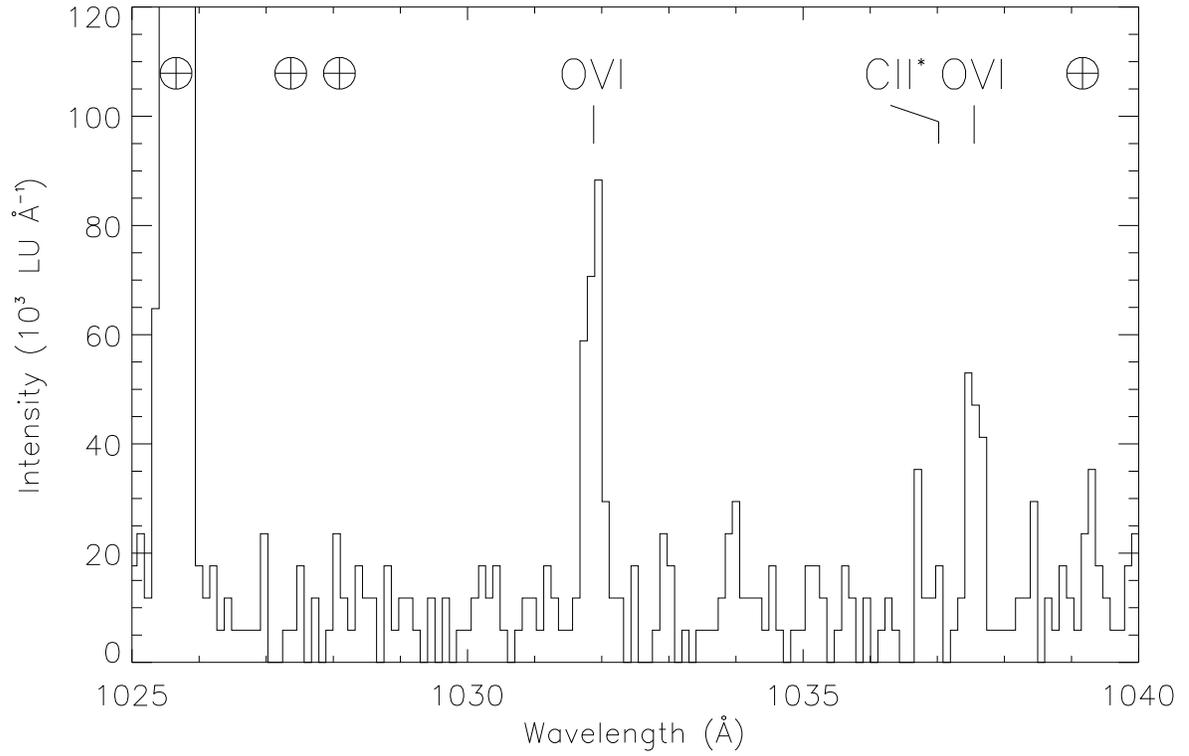}
\caption{\label{spec} Spectrum of sight line S4054403 (map key 3). This sight
line exhibits the strongest \ion{O}{6} emission measured in the nebula around
KPD~0005+5106. The \ion{O}{6} doublet and the \ion{C}{2}$^\ast$ $\lambda$1037
emission line are labeled. The positions of airglow lines are marked with the
Earth symbol.}
\end{figure}

All spectra were fit using the IRAF\footnote{IRAF is distributed by the National
Optical Astronomy Observatories, which are operated by the Association of
Universities for Research in Astronomy, Inc., under cooperative agreement with
the National Science Foundation.} routine SPECFIT \citep{kriss} with
emission-line models consisting of a convolution of a Gaussian and a
106~km~s$^{-1}$ wide flat-top profile representing the LWRS aperture. The
spectrum with the brightest \ion{O}{6} $\lambda$1032 feature (S4054403, map key
3) is shown in Figure \ref{spec}.

The intensities derived from our fits to the \ion{O}{6} emission lines are
listed in Table \ref{meas}. Due to its low signal-to-noise ratio, possible
blending with the \ion{C}{2}$^\ast$ $\lambda$1037 line, and vicinity to the
\ion{O}{1} $\lambda$1039 airglow line, the \ion{O}{6} $\lambda$1038 emission
line was difficult to fit in many cases. To date, the intensities of diffuse
\ion{O}{6} $\lambda$1032 emission detected with {\em FUSE} range from 1700 to
11,000~LU, with an average value of $4700\pm2400$~LU (Otte, Dixon, \& Sankrit
2004, in preparation)\footnote{We note that \ion{O}{6} emission $10-100$ times
brighter is detected in supernova remnant shocks. These, however, are well
identified nebulae and not considered part of the diffuse interstellar medium.}.
With a peak intensity of 25,000~LU, the high-ionization nebula around
KPD~0005+5106 is the brightest region of extended \ion{O}{6} emission yet found.

\section{DISCUSSION}

With a surface temperature $T_{\rm eff}=120,000$~K, KPD~0005+5106 is the hottest
known DO (i.e. helium-rich) WD. The stellar spectrum has been observed at X-ray,
ultraviolet, and optical wavelengths. Model fits yield stellar parameters of
$\log(g)=7$, $E(B-V)=0.13$, and a distance of about 270~pc (Werner, Heber, \&
Fleming 1994; Werner et al. 1996). The star possesses a soft X-ray corona
\citep{odw}, and appears to undergo mass ejection \citep{w96,sion}.

In the $\log(g)-\log(T_{\rm eff})$ diagram, KPD~0005+5106 lies between the
highest-gravity PG~1159 stars known to have PNe and the two DO WD PN candidates
PG~0108+101 and PG~0109+111. The discovery of a PN around the DO WD PG~1034+001
\citep{hew} extends the range of known PNe in the $\log(g)-\log(T_{\rm eff})$
plane along the evolutionary tracks toward lower temperatures and higher surface
gravities ($\log(g)=7.5, T_{\rm eff}=100,000$~K; Werner, Dreizler, \& Wolff
1995). KPD~0005+5106 lies well within this extended parameter space. However,
previous searches for an optical PN around this WD have been unsuccessful
\citep{w97}.

Photoionized \ion{O}{6} is expected from PN with hot central stars
($T_{\rm eff}\geq120,000$~K) and has been detected in a few such objects (Chu,
Gruendl, \& Guerrero 2004). We used the photoionization code CLOUDY to determine
whether photoionization of a putative PN around KPD0005+5106 could explain the
detected \ion{O}{6} emission. For the ionizing source, we used a synthetic
spectrum of the star calculated with the program TLUSTY \citep{hl}, using the
observed stellar abundances and parameters \citep{w96}. The stellar model was
scaled to the flux observed with {\em FUSE}. Simple PN expansion models by
\citet{ost} yield densities of a few cm$^{-3}$ at the distances of the LWRS
aperture positions for a $\sim10,000$~yr old PN. We assumed solar abundances and
a particle density of unity (i.e., $n({\rm H})=1$~cm$^{-3}$) in the absorbing
cloud. With no other free parameters, our model reproduces the observed
\ion{O}{6} flux to within a factor of a few. Based on this agreement between
model predictions and observations, we conclude that the \ion{O}{6} nebula
around KPD~0005+5106 is photoionized. It may be one of a new class of HIPNe that
are faint in standard optical lines.

\section{IMPLICATIONS}

\subsection{The KPD~0005+5106 System}

Our photoionization model predicts a [\ion{Ne}{5}] $\lambda$3426 intensity seven
times as bright as the \ion{O}{6} emission, whereas shock models yield
[\ion{Ne}{5}] intensities $10-100$ times fainter than the \ion{O}{6} emission.
This discriminant has been used to argue that the nebula LMC~62 in the Large
Magellanic Cloud is photoionized \citep{her}. Assuming $E(B-V)=0.13$,
$R_{\rm V}=3.2$, and extrapolating the extinction parameterization of Cardelli,
Clayton, \& Mathis (1989) to the Lyman limit, we find that the [\ion{Ne}{5}]
emission should be about 20 times brighter than the observed \ion{O}{6}
$\lambda$1032 emission around this WD. The photoionization model predicts that
the [\ion{Ne}{5}] emission arises in a more extended zone than the \ion{O}{6}
emission. Though the geometry of the emitting gas is unknown, [\ion{Ne}{5}]
imaging would allow us to confirm the existence of HIPNe around this and other
hot WDs.

Another prediction of a radiation-bounded HIPN is the presence of a low surface
brightness H$\alpha$ and [\ion{O}{3}] $\lambda$5007 nebula with a large angular
diameter as has been observed around PG~1034+001 \citep{hew}, because lower
ionization states are only possible at larger distances from the hot star. The
PN search around KPD~0005+5106 by \citet{w97} likely had too small a field of
view to detect such an extended PN.

\subsection{PN Populations and Evolution}

As the central star of a PN contracts and its temperature rises, its radiation
field becomes harder. Meanwhile, the expansion of the nebula causes its density
to fall. The combination of these effects raises the ionization parameter (the
ratio of photon density to particle density) at the inner face of the nebula,
and with it the ionization state of the gas. This scenario has two implications:
first, that every PN with a central star massive enough to reach a temperature
greater than $\sim120,000$~K will eventually pass through a high-ionization
phase; second, that every hot white dwarf should exhibit a HIPN.

Observations of HIPNe around hot WDs may reveal important information about WD
precursors and their evolution. One unresolved problem, for example, is the
existence of helium-rich, hydrogen-deficient WDs that account for about 20\% of
all known degenerate stars \citep{sion}. With the precursors' abundances
imprinted in the diffuse gas, HIPNe could give clues to the evolutionary process
that results in this bizarre type of WDs.

The photoionization model with unit hydrogen density and solar abundances,
[O/H]$=7.4\times10^{-4}$, predicts that the thickness of the \ion{O}{6} emitting
zone is approximately $5\times10^{18}$~cm. In this region, the fractional
abundance O$^{5+}$/O is about 0.11, which translates to a number density
$N_{{\rm O}^{5+}}\sim8\times10^{-5}$~cm$^{-3}$. The \ion{O}{6} column density is
thus about $4\times10^{14}$~cm$^{-2}$. Depending on the distribution of hot WDs
in the Galactic disk, HIPNe could contribute a significant fraction to the total
\ion{O}{6} column density in the plane.

\acknowledgements

The authors wish to thank J. Dupuis for supplying the stellar spectral model and
R. Benjamin for his insightful comments. The authors also thank J. Herald for
useful discussions. This research is supported by NASA contract NAS5-32985 to
the Johns Hopkins University. This work made use of the NASA Astrophysics Data
System (ADS).

\clearpage

\begin{deluxetable}{ccccccc}
\tablewidth{0pt}
\tablecaption{\label{obs} OBSERVATIONS}
\tablehead{\colhead{} & \colhead{} & \colhead{} & \colhead{} &
\colhead{$\varphi$} & \colhead{Distance} & \colhead{} \\
\colhead{Observation ID} & \colhead{RA\tablenotemark{a}} &
\colhead{DEC\tablenotemark{a}} & \colhead{Observation Date} & \colhead{(deg)} &
\colhead{(pc)} & \colhead{Map Key}}
\startdata
M1070205 & 0 07 56.0 & 51 22 50 & 2001-09-29 & 262.52 & 0.27 & 12 \\
M1070206 & 0 08 06.7 & 51 23 03 & 2001-09-29 & 262.35 & 0.14 & 10 \\
M1070208 & 0 07 59.2 & 51 25 00 & 2002-08-23 & 299.57 & 0.27 & 15 \\
M1070209 & 0 08 08.3 & 51 24 10 & 2002-08-23 & 299.35 & 0.14 & 13 \\
M1070211 & 0 08 00.1 & 51 21 11 & 2002-10-21 & 232.79 & 0.27 & 9 \\
M1070212 & 0 08 08.9 & 51 22 11 & 2002-10-22 & 232.20 & 0.14 & 7 \\
M1070214 & 0 08 00.6 & 51 21 05 & 2002-10-23 & 230.64 & 0.27 & 8 \\
M1070215 & 0 08 09.1 & 51 22 08 & 2002-10-23 & 230.44 & 0.14 & 6 \\
M1070217 & 0 08 22.3 & 51 19 53 & 2002-12-17 & 167.76 & 0.27 & 1 \\
M1070221 & 0 08 09.7 & 51 24 28 & 2003-08-07 & 311.16 & 0.14 & 14 \\
M1070223 & 0 08 17.2 & 51 19 49 & 2003-12-03 & 180.87 & 0.27 & 4 \\
M1070224 & 0 08 17.7 & 51 21 29 & 2003-12-03 & 180.67 & 0.14 & 2 \\
S4053801 & 0 08 04.7 & 51 22 44 & 2001-10-23 & 255.41 & 0.17 & 11 \\
S4053803 & 0 08 11.6 & 51 21 20 & 2003-12-02 & 206.43 & 0.17 & 5 \\
S4054403 & 0 08 15.9 & 51 21 07 & 2002-12-22 & 187.78 & 0.17 & 3 \\
\enddata
\tablenotetext{a}{Coordinates are J2000.0 and refer to the center of the {\em
FUSE} low-resolution aperture. Units of right ascension are hours, minutes,
and seconds, and units of declination are degrees, arcminutes, and arcseconds.}
\end{deluxetable}

\clearpage

\begin{deluxetable}{cccc}
\tablewidth{0pt}
\tablecaption{\label{meas} \ion{O}{6} INTENSITIES}
\tablehead{\colhead{} & \colhead{$t$} & \colhead{$I_{1032}$} &
\colhead{$I_{1038}$} \\
\colhead{Map Key} & \colhead{(sec)} & \colhead{(10$^3$ LU)\tablenotemark{a}} &
\colhead{(10$^3$ LU)\tablenotemark{a}}}
\startdata
1 & 3157 & $<8$\tablenotemark{b} & $<8$\tablenotemark{b} \\
2 & 2773 & $25\pm7$ & $12\pm4$ \\
3 & 3250 & $25\pm4$ & $11\pm3$ \\
4 & 2752 & $18\pm8$ & $12\pm9$ \\
5 & 1890 & $18\pm5$ & $15\pm4$ \\
6/7 & 5789 & $19\pm3$ & $17\pm3$ \\
8/9 & 3953 & $18\pm4$ & $8\pm3$ \\
10 & 2483 & $17\pm4$ & $9\pm3$ \\
11 & 11010 & $16.2\pm2.4$ & $11.1\pm2.8$ \\
12 & 2691 & $20\pm4$ & $16\pm6$ \\
13/14 & 6279 & $12\pm4$ & $5.4\pm2.0$ \\
15 & 4214 & $8\pm3$ & $8.0\pm2.4$ \\
\enddata
\tablenotetext{a}{1 LU = 1 photon s$^{-1}$ cm$^{-2}$ sr$^{-1}$; for \ion{O}{6}
$\lambda$1032, 1 LU = $4.5\times10^{-22}$ erg s$^{-1}$ cm$^{-2}$ arcs$^{-2}$.}
\tablenotetext{b}{Value is a 2 $\sigma$ upper limit.}
\end{deluxetable}

\clearpage

\clearpage

\begin{thebibliography}{}
\bibitem[Cardelli et al.(1989)]{ccm} Cardelli, J. A., Clayton, G. C., \& Mathis,
J. S. 1989, \apj, 345, 245
\bibitem[Chu et al.(2004)]{chu} Chu, Y.-H., Gruendl, R. A., \& Guerrero, M. A.
2004, in ASP Conf. Ser. 313, Asymmetric Planetary Nebulae III, ed. M. Meixner,
J. Kastner, B. Balick, \& N. Soker (San Francisco: ASP), in press
\bibitem[Herald \& Bianchi(2004)]{her} Herald, J. E., \& Bianchi, L. 2004, \apj,
submitted
\bibitem[Hewett et al.(2003)]{hew} Hewett, P. C., Irwin, M. J., Skillman, E. D.,
Foltz, C. B., Willis, J. P., Warren, S. J., \& Walton, N. A. 2003, \apjl, 599,
L37
\bibitem[Hubeny \& Lanz(1995)]{hl} Hubeny, I., \& Lanz, T. 1995, \apj, 439, 875
\bibitem[Kriss(1994)]{kriss} Kriss, G. A. 1994, in ASP Conf. Ser. 61,
Astronomical Data Analysis Software and Systems III, ed. D. R. Crabtree, R. J.
Hanisch, \& J. Barnes (San Francisco: ASP), 437
\bibitem[Kwitter et al.(1989)]{kw} Kwitter, K. B., Massey, P., Congdon, C. W.,
\& Pasachoff, J. M. 1989, \aj, 97, 1423
\bibitem[Moos et al.(2000)]{moos} Moos, H. W., et al. 2000, \apjl, 538, L1
\bibitem[Napiwotzki(1999)]{nap} Napiwotzki, R. 1999, \aap, 350, 101
\bibitem[O'Dwyer et al.(2003)]{odw} O'Dwyer, I. J., Chu, Y.-H., Gruendl, R. A.,
Guerrero, M. A., \& Webbink, R. F. 2003, \aj, 125, 2239
\bibitem[Osterbrock(1989)]{ost} Osterbrock, D. E. 1989, Astrophysics of Gaseous
Nebulae and Active Galactic Nuclei (Mill Valley: University Science Books)
\bibitem[Rauch(1999)]{rau} Rauch, T. 1999, A\&AS, 135, 487
\bibitem[Sahnow et al.(2000)]{sahn} Sahnow, D. J., et al. 2000, \apjl, 538, L7
\bibitem[Sion et al.(1997)]{sion} Sion, E. M., Holberg, J. B., Barstow, M. A.,
\& Scheible, M. P. 1997, \aj, 113, 364
\bibitem[Werner et al.(1997)]{w97} Werner, K., Bagschik, K., Rauch, T., \&
Napiwotzki, R. 1997, \aap, 327, 721
\bibitem[Werner et al.(1996)]{w96} Werner, K., Dreizler, S., Heber, U., Rauch,
T., Fleming, T. A., Sion, E. M., \& Vauclair, G. 1996, \aap, 307, 860
\bibitem[Werner et al.(1995)]{w95} Werner, K., Dreizler, S., \& Wolff, B. 1995,
\aap, 298, 567
\bibitem[Werner et al.(1994)]{w94} Werner, K., Heber, U., \& Fleming, T. 1994,
\aap, 284, 907
\end{thebibliography}
\end{document}